\documentclass[trackchanges]{aastex7}

\begin{document}

\title{Eclipsing Stellar Flare on the Demon Star Algol Binary System \\
Observed during the MAXI-NICER Follow-up Campaign in 2018}

\author
{Kazuya Nakayama}
\affiliation{Department of Physics, Kyoto University, Kitashirakawa Oiwake, Sakyo, Kyoto 606-8502, Japan}
\email[show]{nakayama.kazuya.72r@st.kyoto-u.ac.jp}  

\author[0000-0002-0207-9010
]{Wataru Buz Iwakiri} 
\affiliation{International Center for Hadron Astrophysics, Chiba University, Inage-ku, Chiba, 263-8522, Japan}
\email{iwakiri@chiba-u.jp}

\author[0000-0003-1244-3100
]{Teruaki Enoto}
\affiliation{Department of Physics, Kyoto University, Kitashirakawa Oiwake, Sakyo, Kyoto 606-8502, Japan}
\affiliation{RIKEN Center for Advanced Photonics (RAP), 2-1 Hirosawa, Wako, Saitama 351-0198, Japan}
\email{enoto.teruaki.2w@kyoto-u.ac.jp}

\author[0000-0003-3085-304X]{Shun Inoue}
\affiliation{Department of Physics, Kyoto University, Kitashirakawa Oiwake, Sakyo, Kyoto 606-8502, Japan}
\email{inoue.shun.57c@st.kyoto-u.ac.jp}

\author[0000-0002-0412-0849
]{Yuta Notsu}
\affiliation{Laboratory for Atmospheric and Space Physics, University of Colorado Boulder, 3665 Discovery Drive, Boulder, CO 80303, USA}
\affiliation{National Solar Observatory, 3665 Discovery Drive, Boulder, CO 80303, USA}
\email{yuta.notsu@colorado.edu}

\author[0000-0001-7115-2819]{Keith Gendreau} 
\affiliation{Astrophysics Science Division, NASA’s Goddard Space Flight Center, 8800 Greenbelt Road, Greenbelt, MD 20771, USA}
\email{keith.c.gendreau@nasa.gov}

\author{Zaven Arzoumanian} 
\affiliation{Astrophysics Science Division, NASA’s Goddard Space Flight Center, 8800 Greenbelt Road, Greenbelt, MD 20771, USA}
\email{zaven.arzoumanian-1@nasa.gov}

\author[0000-0001-7515-2779]{Kenji Hamaguchi}
\affiliation{CRESST II and X-ray Astrophysics Laboratory NASA/GSFC,
Greenbelt, MD 20771, USA}
\affiliation{Department of Physics, University of Maryland, Baltimore County,
1000 Hilltop Circle, Baltimore, MD 21250, USA}
\email{kenjih@umbc.edu}

\author[]{Tatehiro Mihara}
\affiliation{RIKEN Pioneering Research Institute, Wako, Saitama 351-0198, Japan}
\email{tmihara@riken.jp}




\begin{abstract}
Algol is a well-known eclipsing binary hosting an active and variable star that exhibits frequent stellar flares. 
Here, we report our pre-planned and coordinated rapid X-ray follow-up observations of an eclipsing flare on Algol. The Monitor of All-sky X-ray Image (MAXI) detected a flare on Algol at 05:52 UT on 2018 July 4. Subsequently, we carried out a prompt X-ray monitoring with the Neutron star Interior Composition Explorer (NICER) starting at 19:45 UT on the same day, and the observation ended at 06:02 UT on 2018 July 6. During the decaying phase of the flare, we successfully detected a 5.8-hour-long eclipse, corresponding to the secondary eclipse in which Algol A blocks the line of sight to Algol B. 
During the eclipse, the 2--10 keV X-ray flux is decreased to 20\% level from $1.9\times10^{-10}~ \mathrm{erg~cm^{-2}~s^{-1} }$ to $4.5\times10^{-11}~ \mathrm{erg~cm^{-2}~s^{-1} }$. 
We found a configuration of the flare size and location to explain the X-ray observations; e.g., the flare occurred at the latitude 45°S of the Algol B surface with a flare height of $1.9\times10^{11}~\mathrm{cm}$, corresponding to 0.8 times the stellar radius of Algol B, giving 80\% obscuration of the flare loop by Algol A. 
The apparent absorption increase before the eclipse might originate from coronal mass ejection (CME) in the line of sight ejected during the flare. 
\end{abstract}

\keywords{\uat{X-ray astronomy}{1810} --- \uat{Algol variable stars}{24} --- \uat{Eclipsing binary stars}{444} --- \uat{Stellar activity}{1580} --- \uat{Stellar astronomy}{1583} --- \uat{Stellar flares}{1603}}


\section{Introduction}\label{intro}
Solar flares are sudden increases of brightness caused by an impulsive release of magnetic energy in the solar atmosphere (e.g., \citealt{key-19} for a review).
Recent studies have revealed a variety of stellar flares, occurring over a very wide range of timescales, from a few seconds to several days, observed across the broad spectral range from X-rays to radio waves (\citealt{key-18} for a review).
In the standard model of solar flares \citep{key-19,key-18}, energy is transported into the lower atmosphere, and the heated chromospheric material is ``evaporated'' into the corona, resulting in strong X-ray emission from coronal flare loops \citep{key-30,key-20}.
A similar process to the ``standard'' model of solar flares is widely discussed for X-ray emissions of stellar flares (e.g., \citealt{key-21,key-29,key-5}).
However, questions remain as to whether the multi-wavelength responses of all stellar flares follow the same process as the standard model of solar flares (e.g., \citealt{key-22,key-23,key-18}, and references therein). 
In particular, since stellar flares cannot be spatially resolved by X-ray imaging observations, it is difficult to obtain information about the flare size and location, complicating studies of the emission and occurrence mechanisms of stellar flares.

An exception to this difficulty is flares on eclipsing binaries, which can provide new insights into such information through analysis of the eclipsing X-ray light curves of stellar flares. 
One good example is the stellar flares of Algol (Beta Per), which is a well-known eclipsing variable star. In ancient times, Algol was the star that represented the head of the monster Medusa, held by Perseus, and was feared as the ``demon star'' due to its optical variability \citep{key-36}. 
Algol is a binary system hosting B8~V-type Algol A and K0~IV-type Algol B, whose spectral types are summarized in Table 1 of  \citet{key-2}, and the basic system parameters in  \citet{key-1} and \citet{key-10}\footnote{Algol is a triple star system in a strict definition.}.
The less massive subgiant star Algol B is considered to have a strong magnetic field and generate frequent flares (e.g., \citealt{key-11}).
The orbital period of the binary, 2.867~days, and its orbital ephemeris are reported by Mt. Suhora Astronomical Observatory (TIDAK; \citealt{key-4}). 
The basic stellar parameters used in this study are summarized in Table \ref{star_param}. 

\begin{deluxetable*}{cccc}
\tablewidth{0pt}
\tablecaption{Stellar and orbital parameters of the Algol binary system. \label{star_param}}
\tablehead{
\colhead{Parameter} & \colhead{Algol A} & \colhead{Algol B} & \colhead{Reference}
}
\startdata
      Radius ($R_{\odot}$) & $2.73\pm0.20$ & $3.48\pm0.28$ & \cite{key-1} \\
      Mass ($M_{\odot}$) & $3.17\pm0.21$ & $0.70\pm0.08$  & \cite{key-1} \\
      Temperature (K) & $13,000$ & $4,500$  & \cite{key-10}\\
      Spectral type & B8 V & K0 IV & \cite{key-2} \\
      Distance (pc) & \multicolumn{2}{c}{$28.8\pm0.5$}  & \cite{key-15} \\
      \hline
      Epoch (MJD) & \multicolumn{2}{c}{$40952.9657$} & \begin{tabular}{c} TIDAK; \cite{key-4} \end{tabular} \\
      \begin{tabular}{c} Period; $P_\mathrm{{orb}}$ (days) \end{tabular} & \multicolumn{2}{c}{$2.8673075$} & \begin{tabular}{c} TIDAK; \cite{key-4} \end{tabular} \\
      \begin{tabular}{c} Semi-major axis (mas) \end{tabular} & \multicolumn{2}{c}{$2.15\pm0.05$}  & \cite{key-1} \\
      Inclination (deg)& \multicolumn{2}{c}{$98.70\pm0.65$}  & \cite{key-1} \\
      \hline
\enddata
\end{deluxetable*}

Algol is a strong source of X-ray emission, exhibiting several stellar flares. 
On 1983 August 18, a flare was detected with EXOSAT with a peak X-ray luminosity $1.4\times10^{31}~\mathrm{erg~s^{-1}}$ in 0.1--10.0~keV and e-folding decay time scale of 7~ks \citep{key-6, key-7}.
Another flare was observed with ROSAT on 1992 August 20 with peak luminosity $1.6\times10^{32}~\mathrm{erg~s^{-1}}$ in 0.1--2.4~keV and e-folding time of 30.4~ks \citep{key-17}. 
The mean occorrence frequency of large X-ray flares ($>10^{31}~\mathrm{erg~s^{-1}}$) on Algol is estimated to be $0.6/P_\mathrm{{orb}}\simeq0.2$ flares per day. 
Eclipses during flares on Algol have been observed twice in the past: one was reported on 1997 August 30 with BeppoSAX, where the flare was thought to occur in the southern polar region with the flare height $H\leq0.6R_{B}$ where $R_{B}$ is the stellar radius of Algol B \citep{key-8, key-11}. 
The other eclipsing flare was observed with XMM-Newton on 2002 February 12, where the flare of $H\simeq0.1R_{B}$ occurred in a part of the northern hemisphere that is not 
a part of the polar region \citep{key-9}. Both of these eclipsing flares were serendipitously detected without any follow-up observation in response to an alert of a flare. 
In the present time-domain astronomy era, detecting stellar flares with wide-field instruments can send an alert to conduct quick follow-up X-ray pointing observations, which is highly effective for ensuring successful eclipsing flare observations.

The Monitor of All-sky X-ray Image (MAXI; \citealt{key-3}) scans the entire sky approximately every 92 minutes in X-rays and has detected many stellar flares (e.g., \citealt{key-24}). Catching the decaying phase of these flares requires prompt X-ray follow-ups, typically within  1-3 hours for M-type dwarfs and 1-27 hours for RS-CVn-type stars \citep{key-24}. Thus, in 2017, the MAXI and  Neutron star Interior Composition Explorer (NICER; \citealt{key-14}) teams started a new follow-up observation system that allows NICER to perform a prompt pointing observation of transient events detected by MAXI. This campaign is named MANGA (MAXI And NICER Ground Alert). The downlinked MAXI data is checked in the ground system by two transient searching algorithms with the unbiased nova search \citep{key-28} and the optimized search for predefined X-ray bursters or stellar flare sources (called MANGA trigger). When either of the two searching algorithms finds a transient, after a human final decision, a follow-up request is sent to NICER operations team over the internet. This MANGA system made several successful stellar flare detections \citep{key-25,key-26,key-27}. Here, we report our successful observation of the eclipsing X-ray flare on Algol using the MANGA system.

\section{Observations and Data Reduction}\label{sec:2}

MAXI detected a stellar flare from Algol on 2018 July 4 at 05:52 UT with a flux of about 300 mCrab \citep{key-35}. The MANGA collaboration conducted follow-up observations with NICER starting on 2018 July 4 at 19:45 UT. 
Figure~\ref{lc} shows the NICER count rate during this Algol flare event, adding the MAXI monitoring data normalized to the NICER count rate. We use PIMMS \citep{key-15} to convert the MAXI 2.0$-$5.0 keV count rate into the NICER 2.0$-$5.0 keV count rate. After the initial scan at the flare detection, MAXI covered the flare peak at the second scan, whereas NICER started the follow-up observation 14 hours after the flare peak. The NICER observation continued for 34 hours, and 
the X-ray flux in the 2.0--10.0~keV was $\mathrm{(3.9\pm1.1)\times10^{-9}~erg~cm^{-2}~s^{-1}}$ at the peak and $\mathrm{(1.23\pm0.02)\times10^{-10}~erg~cm^{-2}~s^{-1}}$ at the last snapshot of the NICER observation, which is by $\mathrm{1.12\times10^{-10}~erg~cm^{-2}~s^{-1}}$ higher than the quiescent flux $\mathrm{(1.14\pm0.06)\times10^{-11}~erg~cm^{-2}~s^{-1}}$ observed on 2018 August 11 at 18:20--20:00 UT (ObsID 1200260104, exposure time 838 sec).

We downloaded the NICER data of ObsIDs~1200260101--1200260106 from the HEASARC archive. We employed the standard calibration and screening procedures as described in \citet{key-27}. First, we used \texttt{nicerl2} in HEASoft ver. 6.33.2 and the calibration database (\texttt{CALDB}) version \texttt{xti20240206} to filter and calibrate raw data with no additional options. Then, we extracted X-ray light curves from the filtered events with xselect and generated ObsID-averaged on-source and background spectra with \texttt{nicerl3-spect} and the SCORPEON background model \citep{key-13}. We also extracted time-resolved spectra of ObsIDs 1200260101--1200260103 with the \texttt{nimaketime}, \texttt{niextract-events}, and \texttt{nicerl3-spect} commands. For the observation during the quiescent phase, we downloaded ObsID 1200260104 and performed the same procedures. We used Xspec ver.12.14.0h \citep{key-12} for our spectral analysis.
We also downloaded the MAXI data from the MAXI on-demand process \citep{key-16}. For the observations from MJD 58303.15 to 58303.80, we obtained the MAXI data at each scan.
In addition, we downloaded the TESS \citep{key-32} data from the Barbara A. Mikulski Archive for Space Telescopes (MAST) archive at the Space Telescope Science Institute (STScI). The TESS data from 2019 November 3 was also used with a 2-min time cadence, which is the closest observation to the MAXI and NICER campaign for this flare.

\section{Analysis and Results}\label{result}

\begin{figure}
 \begin{center}
  \includegraphics[width=16cm]{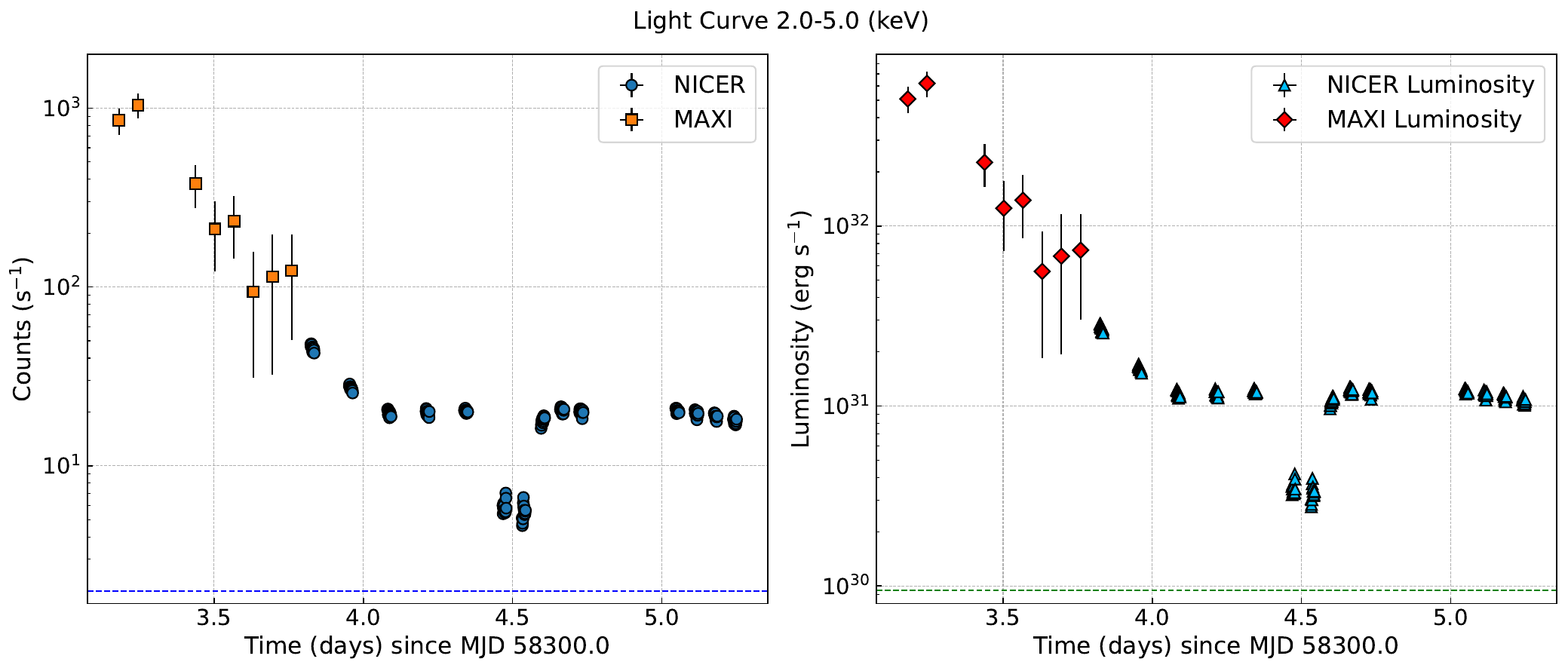} 
 \end{center}
\caption{NICER (blue circles, skyblue triangles) and MAXI (orange squares, red diamonds) count rates and luminosity in the 2.0--5.0~keV band of Algol, which are binned at 64~s and 211~s (each MAXI scan) in the left and right panels, respectively. 
The MAXI data are added after being converted to the corresponding NICER count rate by PIMMS, assuming that the MAXI spectrum can be represented by a power law and the temperture was 4~keV.
The time zero of MJD 58300.0 corresponds to 2018 July 1, 00:00 UT. 
The blue dotted line and green dotted line show the 2.0--5.0 keV NICER count rate and luminosity in the quiescent state on 2018 August 11, $2.0~\mathrm{counts~s^{-1}}$ and $9.5\times10^{29}~\mathrm{erg~s^{-1}}$ in the left and right panels, respectively.
}\label{lc}
\end{figure}

As shown in Figure~\ref{lc}, X-ray dimming was observed during the decaying phase of the flare between 
MJD 58304.4 and 58304.6, 
suggesting that the flare was obscured partially or completely by either of the binary stars. 
During the dimming, the NICER 2.0--5.0~keV count rate decreased from $20~\mathrm{counts~s^{-1}}$ to $6.5~\mathrm{counts~s^{-1}}$. Taking into account the NICER count rate at the quiescent state ($2.0~\mathrm{counts~s^{-1}}$) obtained in 2018 (ObsID 1200260104), the NICER rate decreased by approximately 80\% during the dimming.
In order to determine the start and end times of this dimming, we fit the NICER light curve using two error functions,
$f(t) = A_1 (1 + \mathrm{erf}\{(t - \mu_1)/\sigma_1\}) + A_2 (1 - \mathrm{erf}\{(t - \mu_2)/\sigma_2\}) + const$, where $A_1$ and $A_2$ are the amplitudes of the rising and falling edges, $\mu_1$ and $\mu_2$ represent the center positions of these transitions, $\sigma_1$, and $\sigma_2$ are the corresponding widths controlling the steepness, respectively.
Fitting the data by this function, we evaluated the dimming start and end ranges between $\mu_1-1.5~\sigma_1$ and $\mu_2+1.5~\sigma_2$, as MJD 58304.34--58304.47 and MJD 58304.60--58304.67, respectively. 

To identify the binary orbital phase ($\phi_{\mathrm{orb}}$) at the dimming, we compared the corresponding orbital phase with optical observations.
In the Algol-type binary system, there is a known discrepancy of the time of the minimum in the light curve calculated between the reference epoch with the binary period (C) and the actually observed time (O). This difference is called ``O-C'' due to physical effects, for example, mass transfer from a companion star and N-body gravitational interactions.
Therefore, we need this correction before computing the phase. Previous studies \citep{key-8, key-9} did not take the O–C into account, whereas we determined the orbital phase more accurately using the O-C information. The time of the primary optical brightness minimum (main eclipse) is set as $\phi_{\mathrm{orb}}=0$.
The O-C data of the Algol system provided by Mt. Suhora Astronomical Observatory \citep{key-4} 
reported 0.12~day when the flare was observed on 2018 July 4. Here, ``C'' is calculated using $\mathrm{JD}~2440953.4657 + 2.8673075~\mathrm{E}$, where E is an arbitrary integer and C is MJD 58303.16. Therefore, MJD 58303.28 is set as $\phi_{\mathrm{orb}}=0$ in the present NICER data. Assuming this ephemeris, the observed dimming occured at $\phi_{\mathrm{orb}}=0.468$, corresponding to the second eclipse when Algol A blocked Algol B, suggesting that the flare occurred on Algol B. 

\begin{figure}
 \begin{center}
  \includegraphics[width=16cm]{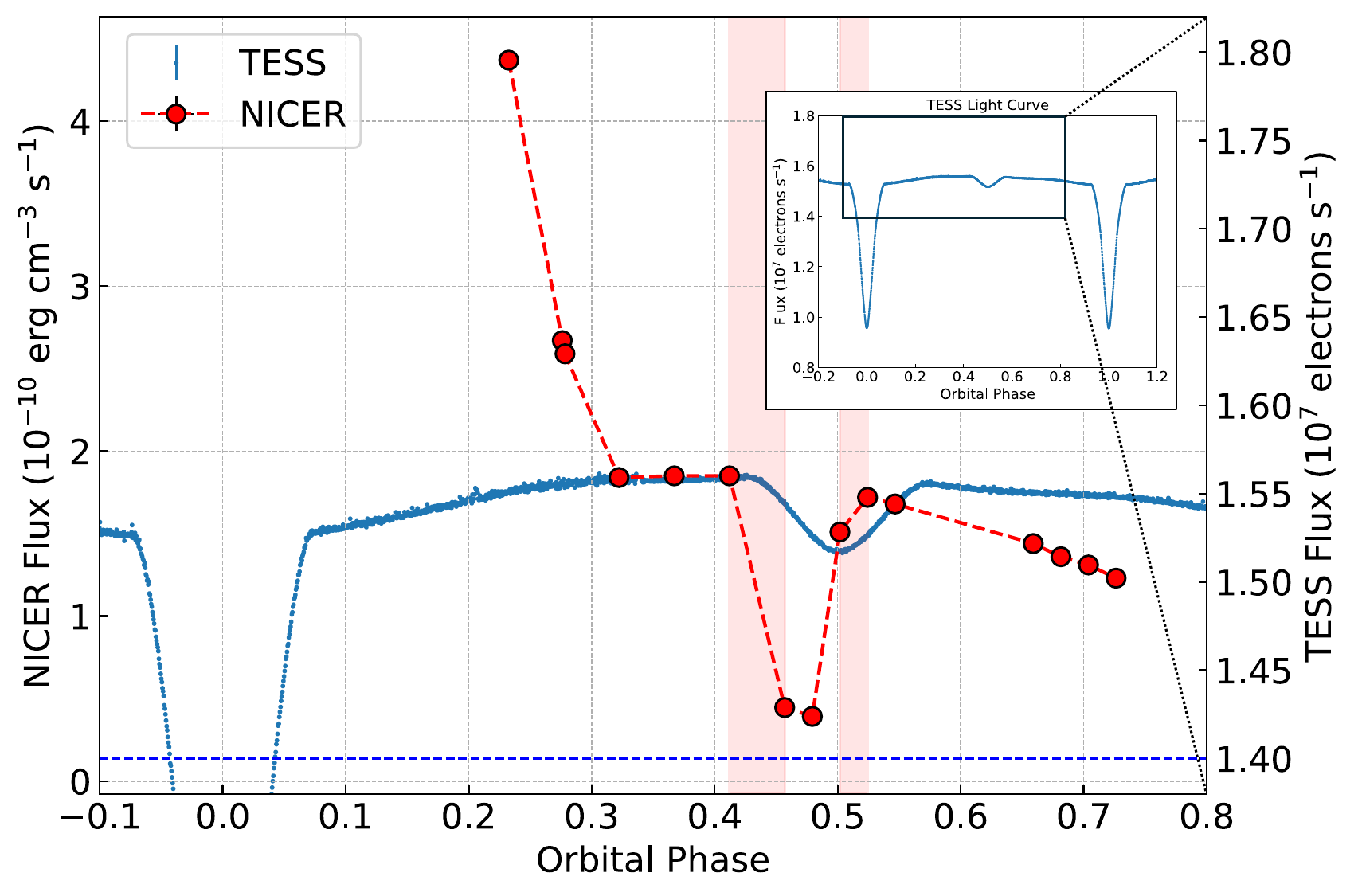} 
 \end{center}
\caption{Optical observation of Algol by TESS on MJD 58793.2--58795.8 (Photometer, 600--1000~nm wavelength, blue dot) folded at the orbital period of Algol ($P_\mathrm{{orb}}$ in Table \ref{star_param}). The inset figure shows this TESS observation covering the entire orbital period. The 2.0--10.0~keV absorbed flux from the NICER flare observations is overlaid. The red shaded areas show the eclipse start and end times of X-rays. The blue horizontal dotted line shows the 2.0-10.0~keV NICER flux at the quiescent level of $1.4\times10^{-11}~\mathrm{erg~cm^{-2}~s^{-1}}$. It should be noted that the TESS and NICER observations were not conducted simultaneously.
}\label{flux_fit}
\end{figure}

Additionally, to compare with available optical data, the TESS observation on MJD 58793.2--58795.8 (one year later, 2019 November 6--8) is converted into the orbital phase. 
The time of minimum brightness in the TESS data is determined by fitting the light curve with the aforementioned error function and identifying the midpoint between $\mu_1$ and $\mu_2$.
This is intended for comparison purposes, and it should be noted that the optical observations were not conducted simultaneously with X-rays.
Figure~\ref{flux_fit} shows the TESS and NICER data folded at $P_\mathrm{{orb}}=2.8673075$ days. The observed dimming is confirmed to be consistent with the second eclipse and the X-ray eclipse starts at an earlier phase than the visible light eclipse by 0.086 phase.

\begin{figure}
 \begin{center}
  \includegraphics[width=16cm]{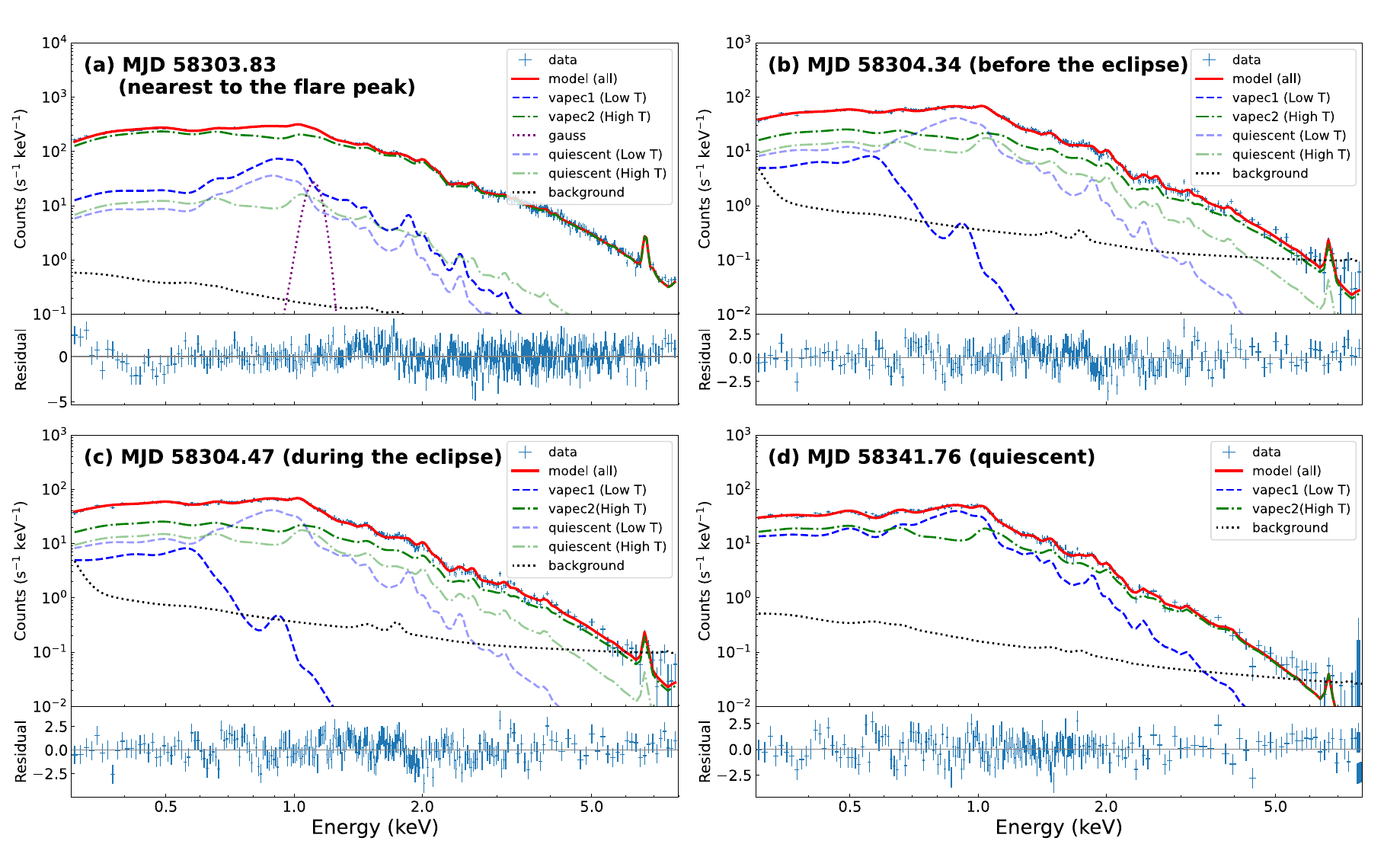} 
 \end{center}
\caption{Background-subtraced Algol flare spectra in the 0.3--8.0~keV band for the first NICER observation on MJD 58303.83 (ObsID 1200260101, exposure time 1068~sec, panel~a), before the eclipse on MJD 58304.34 (ObsID 1200260102, exposure time 1104~sec, panel~b), during the eclipse on MJD 58304.47 (ObsID 1200260102, exposure time 1104~sec, panel~c), and quiescent on MJD 58341.76 (ObsID 1200260104, exposure time 1024~sec, panel~d). The flare spectra are fitted with the $\mathrm{tbabs(vapec+vapec+vapec+vapec)}$ model. 
The quiescent spectrum is fitted with the two-temperature CIE plasma model, and its best-fit result is included in each flare fit with fixed parameters.
The additional two collisionally-ionized components of the flare are fitted, with the best-fit models shown by the solid red line, composed of low-temperature (blue dashed), high-temperature (green dashed-dotted), and an added Gaussian at the Fe-L line (purple dotted). The residuals, $\mathrm{(data-model)/error}$, are shown in the bottom panels. The expected NICER background spectra (black dotted) are calculated with the SCORPEON model and shown for comparison. 
}\label{spec}
\end{figure}

Next, we fit the spectra and the results are shown in Figure~\ref{spec}.
We firstly fit the NICER 0.3--8.0 keV quiescent spectra (ObsID 1200260104) with a collisionally ionized equilibrium (CIE) model composed of two temperature components ($2\times\mathrm{vapec}$ in Xspec). The quiescent spectrum is shown in Figure~\ref{spec}d. The temperature ($T$) and emission measure (\textit{EM}) of the high-temperature component are $T_\mathrm{H}=(2.3\pm0.1)\times10^7~\mathrm{K}$ and $EM_\mathrm{H}=(2.0\pm0.1)\times10^{53}~\mathrm{cm}^{-3}$, and those of the low-temperature component are $T_\mathrm{L}=(9.6\pm0.2)\times10^6~\mathrm{K}$ and $EM_\mathrm{L}=(1.2\pm0.1)\times10^{53}~\mathrm{cm}^{-3}$, respectively.
We secondly fit the flare spectra (ObsIDs 1200260101--1200260103) with an interstellar medium (ISM) absorption model (tbabs) and CIE model composed of four temperature components $4\times\mathrm{vapec}$. Of these, two vapec components we fixed at the temperature and normalization values corresponding to the quiescent phase. Additionally, tbabs was not used for the last four points of the observation (after MJD 58305.0, ObsID 1200260103) since no absorption is required. Figure~\ref{spec}a-c shows the best-fit spectrum of the first NICER observation, before the eclipse, and during the eclipse. In the first NICER spectrum, a Gaussian component centered at 1.1 keV was added to accommodate the Fe L-line, which arises in transitions from the L-shell of iron atoms (Figure~\ref{spec}a). The obtained best-fit spectral parameters, energy flux ($F$), temperature, and \textit{EM} are summarized in Table \ref{fit_param}.

\begin{deluxetable*}{ccccc}
\tablewidth{0pt}
\tablecaption{The best-fit spectral parameters of the Algol flare. \label{fit_param}}
\tablehead{
\colhead{Parameter} & \colhead{First Observation} & \colhead{Before the Eclipse} & \colhead{During the Eclipse} & \colhead{Quiescent}
}
\startdata
      $F_\mathrm{S}$\tablenotemark{a} ($\mathrm{10^{-11}~erg~cm^{-2}~s^{-1}}$)& $41\pm0.5$ & $19\pm0.1$ & $7.9\pm0.1$ & $4.2\pm0.1$ \\
      $F_\mathrm{H}$\tablenotemark{b} ($\mathrm{10^{-11}~erg~cm^{-2}~s^{-1}}$) & $44\pm0.5$ & $19\pm0.3$ & $4.5\pm0.1$ & $1.1\pm0.1$\\
      \begin{tabular}{cc} $T_\mathrm{L}$ & ($10^6$~K) \\ & (keV) \end{tabular} & \begin{tabular}{c} $10\pm0.4$ \\ $0.90\pm0.04$ \end{tabular} & \begin{tabular}{c}$13\pm0.8$ \\ $1.15\pm0.07$ \end{tabular} & \begin{tabular}{c}$1.6\pm0.2$ \\ $0.14\pm0.02$ \end{tabular} & \begin{tabular}{c}$9.6\pm0.2$ \\ $0.83\pm0.02$ \end{tabular} \\
      \begin{tabular}{cc} $T_\mathrm{H}$ & ($10^6$~K) \\ & (keV) \end{tabular} & \begin{tabular}{c} $48\pm1.2$ \\ $4.15\pm0.11$ \end{tabular} & \begin{tabular}{c} $46\pm2.2$ \\ $3.96\pm0.19$ \end{tabular} & \begin{tabular}{c} $38\pm2.3$ \\ $3.26\pm0.20$ \end{tabular} & \begin{tabular}{c} $23\pm1.0$ \\ $2.01\pm0.09$ \end{tabular}\\
      $EM_\mathrm{L}$($10^{53}~\mathrm{cm^{-3}}$) & $2.7\pm0.5$ & $1.6\pm0.6$ & $0.2\pm0.1$ & $1.2\pm0.1$ \\
      $EM_\mathrm{H}$($10^{53}~\mathrm{cm^{-3}}$) & $45\pm0.4$ & $19\pm0.5$ & $3.9\pm0.1$ & $2.0\pm0.1$\\
      $N_{\mathrm{H}}~(10^{20}~\mathrm{cm^{-2}})$ & $2.57\pm0.10$ & $3.10\pm0.13$ & $1.42\pm0.34$ & - \\
      \hline
\enddata

\tablenotemark{a}The soft band absorbed X-ray flux in the 0.3-2.0 keV band.  

\tablenotemark{b}The hard band absorbed X-ray flux in the 2.0-10.0 keV band.   

\tablenotemark{}The L and H subscripts represent the low and high-temperature components of the CIE model, respectively. The uncertainties are given at the 90\% confidence level.
\end{deluxetable*}

The X-ray peak of the 2.0--5.0~keV light curve (Figure~\ref{lc}) was at 05:52 UT during the MAXI second scan.
The peak flux is $(6.2\pm1.0)\times10^{-9}~\mathrm{erg~cm^{-2}~s^{-1}}$ in the 2.0--5.0~keV, and 
this corresponds to the peak luminosity of 
$(6.20\pm0.98)\times10^{32}~\mathrm{erg~s^{-1}}$ assuming the source distance at 28.8~pc \citep{key-15}. 
When fitting the initial one-day light curve during MJD 58303.0 and 58304.0 with an exponential function, the e-folding time is $18\pm1~\mathrm{ks}$. 

As shown in Figure \ref{lc}, the X-ray light curve exhibits two components: one exponentially decaying component from the flare peak (e-folding time of $18\pm1~\mathrm{ks}$ on MJD 58303.0--58304.0), and the other component after MJD 58304.0 remained nearly constant until the end of the NICER observation. 
For the former component, the flux decreased from $(3.9\pm1.1)\times10^{-9}~\mathrm{erg~cm^{-2}~s^{-1}}$ to $(2.6\pm0.1)\times10^{-10}~\mathrm{erg~cm^{-2}~s^{-1}}$ and the total emitted energy in the 2.0--10.0~keV band was $E=(1.3\pm0.2)\times10^{37}~\mathrm{erg}$ during MJD 58303.0--58304.0. 
For the latter component, the flux decreased from $(1.8\pm0.1)\times10^{-10}~\mathrm{erg~cm^{-2}~s^{-1}}$ to $(1.2\pm0.1)\times10^{-10}~\mathrm{erg~cm^{-2}~s^{-1}}$ and the total emitted energy was $E=(1.6\pm0.1)\times10^{36}~\mathrm{erg}$ in 2.0--10.0~keV during MJD 58304.0--58305.25 (until the last NICER observation). 
The eclipse occurred during the latter component. 
When fitting the latter component using the light curve model of $A\mathrm{exp}(-t/\tau)+const$, where $A$ is the amplitude, $\tau$ the decay time constant, the flux is expected to decrease by 99\% at MJD 58316.1--58319.7 to the quiescent level of $(1.1\pm0.1)\times10^{-11}~\mathrm{erg~cm^{-2}~s^{-1}}$.

\begin{figure}
 \begin{center}
  \includegraphics[width=14cm]{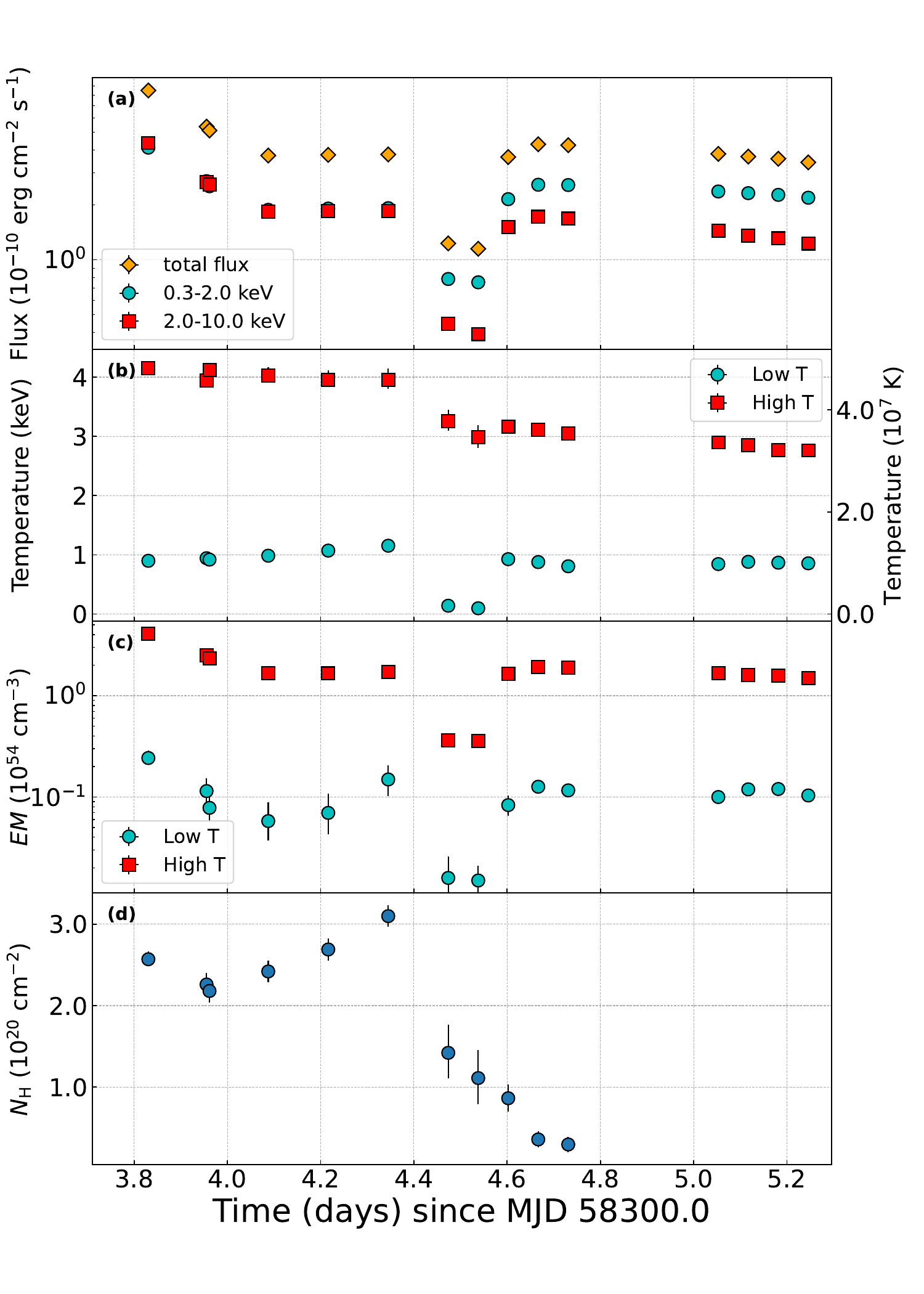} 
 \end{center}
\caption{Time evolutions of the best-fit NICER spectral parameters.~(a)~Absorbed flux in the 0.3--2.0~keV (blue circle), 2.0--10.0~keV (red squere), and total (0.3--10.0~keV, orange diamond) bands.~(b)~Temperatures of the two plasma components.~(c)~Emission Measure (\textit{EM}) calculated from the normalization of the vapec components, i.e., $EM=4\pi d^2k\times10^{14}$ where $d$ is the distance to the source (cm) assuming $d=28.8~\mathrm{pc}=8.9\times10^{19}~\mathrm{cm}$ for Algol and $k={10^{-14}\over{4\pi d^2}}\int n_en_HdV$, where $n_e$ and $n_H$ are the electron and hydrogen densities (cm$^{-3}$), respectively is the normalization derived in Xspec.~(d)~Hydrogen equivalent column density ($N_{\mathrm{H}}$). The last four points are not shown since the spectra are fitted well without the ISM absorption component.
}\label{parameta}
\end{figure}

We summarize the time evolution of the spectral parameters in Figure~\ref{parameta}. The absorbed 2.0--10.0~keV flux decreased faster than that in the 0.3--2.0~keV band (Figure~\ref{parameta}a) especially at the latter part of the NICER observations, and also during the eclipse. The corresponding temperature of the high-temperature component decreased during the eclipse (Figure~\ref{parameta}b). These suggest the hotter region (e.g., flare loop) was obscured. In each of the energy bands, the flux decreased by approximately 80\% during the eclipse. \textit{EM} also decreased from $(19\pm0.5)\times10^{53}~\mathrm{cm}^{-3}$ to $(3.9\pm0.1)\times10^{53}~\mathrm{cm}^{-3}$ in the high-temparature component and from $(1.6\pm0.6)\times10^{53}~\mathrm{cm}^{-3}$ to $(0.2\pm0.1)\times10^{53}~\mathrm{cm}^{-3}$ in the low-temparature component, during the eclipse (Figure~\ref{parameta}c). The hydrogen equivalent column density ($N_{\mathrm{H}}$) increased to $(3.10\pm0.13)\times10^{20}~\mathrm{cm}^{-2}$, which was the maximum value in this NICER observation, before the eclipse and decreased rapidly to $(1.42\pm0.34)\times10^{20}~\mathrm{cm}^{-2}$ during and after it (Figure~\ref{parameta}d). 

\section{Discussion}
\begin{figure}
 \begin{center}
  \includegraphics[width=16cm]{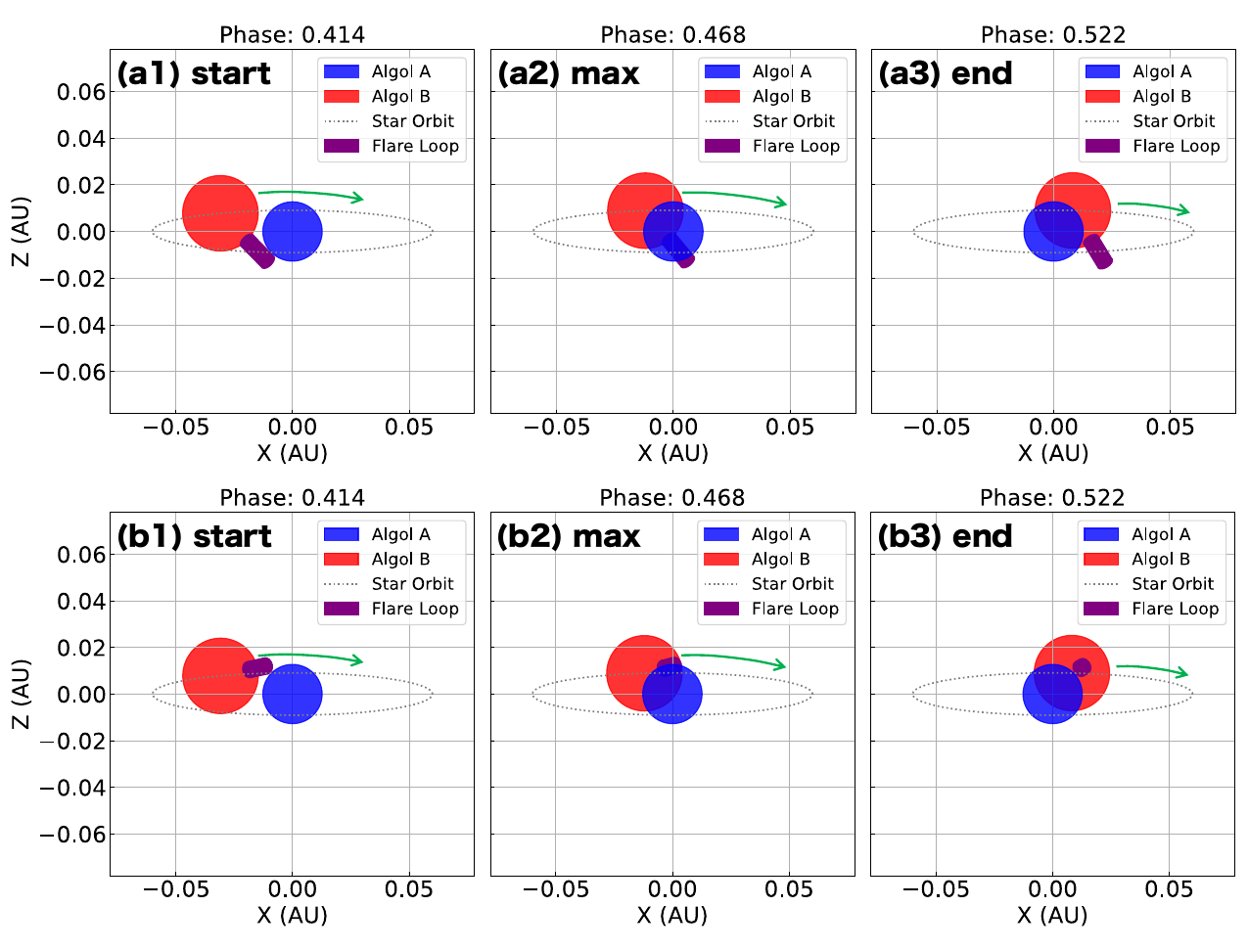} 
 \end{center}
\caption{The observed configurations around and during the second eclipse shown in the line of sight of Algol and two possible examples of the flare. The upper panels a1--a3 show the flare occurring in the mid-latitude region of the southern hemisphere of Algol B (Example A), whereas the lower panels b1--b3 show the flare in the low-latitude region of the northern hemisphere (Example B). The green arrows indicate the moving orbital direction of Algol B. As the orbital motion progresses, the flare loop moves to the near side due to the tidally-locked rotation of Algol B. Panels a1 and b1 correspond to the eclipse start ($\phi_{\mathrm{orb}}=0.414$) when the flare loop begins to be obscured by Algol A. Panels a2 and b2 are at the center of the eclipse ($\phi_{\mathrm{orb}}=0.468$) when 80\% of the flare loop is obscured. Panels a3 and b3 are the end of the eclipse ($\phi_{\mathrm{orb}}=0.522$), when the flare loop completely moves out of the region blocked by Algol A.
}\label{flare}
\end{figure}

Here, we estimate the location and height of the MAXI-NICER flare to explain the observations described in Section \ref{result}. Since the Algol binary system has a short orbital period of $P_\mathrm{{orb}}\simeq2.867$ d, and the two stars are thought to be tidally locked, the stellar rotation period of Algol B is assumed to be the same as $P_\mathrm{{orb}}$. Then, we approximated the flare loop as a cylinder structure attached and perpendicular to the stellar surface, the interior of which is filled with a uniform CIE plasma. 
Figure~\ref{flare} shows the two examples identified in our numerical study. These examples are suggested based on the following observational requirements. 
\begin{enumerate}
    \item The flare loop begins to be obscured by Algol A at the start time of the eclipse at $\phi_{\mathrm{orb}}=0.414$ (Figure~\ref{flare} a1, b1).
    \item Based on the flux decrease, 80\% of the flare loop is obscured at the mid-eclipse at $\phi_{\mathrm{orb}}=0.468$ (Figure~\ref{flare} a2, b2), when assuming uniform X-ray emission from the flare region.
    \item The flare loop completely moves out of the region blocked by Algol A at the end of the eclipse at $\phi_{\mathrm{orb}}=0.522$ (Figure~\ref{flare} a3, b3).
\end{enumerate}
One example of the flare configuration that can explain the observations is shown in Figure~\ref{flare} panels a1-a3. The flare height is assumed to be  $H=1.9\times10^{11}~\mathrm{cm}=0.8R_{B}$ and the location of the flare is on the stellar surface of Algol B at 45°S latitude in the south hemisphere (Example A). Another example to satisfy the requirements is shown in Figure~\ref{flare} panels b1-b3, where $H=1.5\times10^{11}~\mathrm{cm}=0.6R_{B}$ and the location of the flare is at 15°N (Example B). 
Both Example A and B satisfy the two conditions above for reproducing the eclipse. However, Example B does not reproduce the flare light curve before the eclipse. Since the flare occurs at mid-latitudes at Example B, the flare loop was located behind Algol B in the early phase of the flare (before MJD 58303.5), and was not visible. This is inconsistent with the observed curve. Moreover, flares occurring in the region between the two flare examples (between 45°S and 15°N) are entirely blocked by Algol A at the time of maximum eclipse, and thus cannot perfectly reproduce the light curve. Therefore, the configuration of Example A is preferred to explain the present event.

As an independent evaluation, we also estimated the flare loop length using our X-ray spectroscopy based on the equation 
\begin{eqnarray}
 L = 10^{9}\left(\frac{EM}{10^{48}~\mathrm{cm^{-3}}}\right)^{3/5}\left(\frac{n_{0}}{10^{9}~\mathrm{cm^{-3}}}\right)^{-2/5}\left(\frac{T}{10^{7}~\mathrm{K}}\right)^{-8/5}~\mathrm{cm}
 \label{length}
\end{eqnarray}
from \citet{key-5}, where $L$, $EM$, $n_0$, and $T$ are the flare loop length, the peak emission measure, the preflare electron density, and the peak temperature, respectively. This equation is commonly used when calculating $L$ from the X-ray spectroscopy of stellar flares \citep{key-44, key-45, key-46}. 
The flare peak parameters at the MAXI scan are  $EM=3.7\times10^{55}~\mathrm{cm^{-3}}$ and $T=6.0~\mathrm{keV}=7.0\times10^{7}~\mathrm{K}$ ($T$ is fixed to a constant value when calculating the $EM$ in Section~\ref{result}).
Since we do not know $n_0$ from the observations, we assumed a typical range of $n_{0}=10^{10-12}~\mathrm{cm^{-3}}$ from the X-ray grating observations in \cite{key-40}. 
As a result, we estimated $L=(1.0-6.2)\times10^{11}~\mathrm{cm}=0.4-2.6R_{B}$. If $H$ is thought to be half the flare loop length, then $H=(0.5-3.1)\times10^{11}~\mathrm{cm}=0.2-1.3R_{B}$ is derived. 
The loop height from the eclipse geometry (in the above) is in the range estimated here, and this suggests the both estimates can be consistent.

We also conduct a similar calculation using the NICER first observation data (MJD 58303.83; ObsID 1200260101) without assuming the fixed temperature. 
The parameters used are  $EM=4.7\times10^{54}~\mathrm{cm^{-3}}$ and $T=4.6\times10^{7}~\mathrm{K}$. $T$ is calculated as the $EM$-weighted average. 
As a result, we estimated $L=(0.6-3.5)\times10^{11}~\mathrm{cm}=0.2-1.4R_{B}$ and  $H=(0.3-1.7)\times10^{11}~\mathrm{cm}=0.1-0.7R_{B}$. 
This flare height is slightly shorter than the value estimated from the geometry of Example A, but considering that the spectral values used in this estimation were taken slightly after the flare peak, the results show a good agreement.
The equation (\ref{length}) holds when the energy balance between conduction heating and reconnection heating is maintained, we also calculate $L$ using the NICER data before the eclipse (MJD 58304.34; ObsID 1200260102). The used parameters are  $EM=2.0\times10^{54}~\mathrm{cm^{-3}}$ and $T=4.3\times10^{7}~\mathrm{K}$. $T$ is calculated as the $EM$-weighted average. 
As a result, we estimated $L=(0.4-2.3)\times10^{11}~\mathrm{cm}=0.2-1.0R_{B}$ and  $H=(0.2-1.2)\times10^{11}~\mathrm{cm}=0.1-0.5R_{B}$. 
Thus, the flare height derived from the eclipse light curve profile and that derived from the X-ray energy spectrum are consistent with each other.


We compared the present flare with previously reported two eclipsing flares as summarised in Table \ref{pre_flare}.
All three stellar flares in 1997, 2002, and 2018 were estimated to occur on Algol B. The peak luminosity of the present 2018 flare was $7.04^{+0.7}_{-1.9}\times10^{32}~\mathrm{erg~s^{-1}}$ and this was one of the largest X-ray flares ever reported from Algol. This luminosity is comparable to that of the 1997 event  ($3.2\times10^{32}~\mathrm{erg~s^{-1}}$), whereas the 2018 loop size ($0.6R_B$ or $0.8R_B$) is slightly larger than the 1997 size ($0.6R_B$).
The peak luminosity ($6.7\times10^{30}~\mathrm{erg~s^{-1}}$) and flare size ($0.1R_B$) in 2002 are smaller than the present 2018 event.
Thus, our reported 2018 event is the largest, or one of the largest, eclipsing X-ray flares ever observed. 
As a comparison, there were two radio and X-ray observations on Algol in 2008 and 2000, respectively, during the quiescent state without a clear flare and the reported coronal loop heights were the order of a stellar radius both \citep{key-39, key-41}. 
These are slightly larger than the loop size estimated in this study, 
which could originate from the difference in the emitting region of the persistent corona and flares or from long-term variations.

\begin{deluxetable*}{cccccc}
\tablewidth{0pt}
\tablecaption{The flare parameters inferred from the eclipse in Algol. \label{pre_flare}}
\tablehead{
\colhead{Date} & \colhead{Flare height} & \colhead{Location} & \colhead{Peak luminosity $(\mathrm{erg~s^{-1}})$} & \colhead{Peak timing ($\phi_{\mathrm{orb}}$)} & \colhead{Reference}
}
\startdata
      1997/08/30 & $0.6R_B$ & South pole of Algol B & $3.2\times10^{32}$ (0.1--10.0~keV) & 0.1 & \cite{key-8} \\
      2002/02/12 & $0.1R_B$ & Algol B and did not occured in polar region & $6.7\times10^{30}$ (0.2--10.0~keV) & 0.51 & \cite{key-9} \\
      2018/07/04 Example A & $0.8R_B$ & 45°S of Algol B & $7.04^{+0.7}_{-1.9}\times10^{32}$ (0.1-10.0~keV)\tablenotemark{$*$} & 0.027 & - \\
      \hline
\enddata

\tablenotetext{$*$}{The extrapolation from the best-fit model of MAXI observation in the 2.0--10.0~keV.}  

\end{deluxetable*}

Finally, in Figure~\ref{lc}, the NICER count rate before the eclipse did not decrease monotonically, but remained constant at the elapsed day of 0.8--1.1 day (the second component). The spectral parameters also remain constant during this interval, as shown in Figure~\ref{parameta}. This ``plateau'' is an interesting feature, and has been observed several times (ex. \citealt{key-48}). This is rarely seen in M stars compared to RS CVn stars (ex. \citealt{key-23}), but detailed studies have not yet been conducted. A statistical comparison will be necessary in the future. Moreover, in the case of $N_{\mathrm{H}}$, it even starts to increase. When the tbabs component was removed from the best-fit model of the observation where $N_{\mathrm{H}}$ reached its maximum before the eclipse, the chi-square value worsened from 441.2 to 2313.6 at the degree of freedom of 384, indicating that absorption is statistically required.
These facts suggest that another flare could occur during the ongoing flare. The increase in $N_{\mathrm{H}}$ can be interpreted as an effect a coronal mass ejection (CME) due to the second flare, which provides the increase of absorbing material. A similar $N_{\mathrm{H}}$ increases were also observed in the previous flares on 1997 August 3 \citep{key-11} and 1997 August 30 \citep{key-37}, where the possibility of a CME was also suggested. 
\cite{key-38} also reported a potential CME based on the variation of $N_{\mathrm{H}}$.
More studies are needed how we estimate the CME physical parameters from these $N_{\mathrm{H}}$ changes. 
Another hypothesis is that, over time, another X-ray active region might have emerged or rotated into view. However, there is no definitive evidence to distinguish this from a second flare occurring in the same active region.

\section{Summary}
We detected a stellar flare from the binary system Algol at 05:52 UT on 2018 July 4 with MAXI. The flare peak was covered during the second MAXI scan. We carried out follow-up observations with NICER starting 14 hours after the flare peak until 06:02 UT on 2018 July 6. We derived the following conclusions from these observations;

\begin{itemize}
  \item We detected the secondary eclipse in X-ray at 09:30--15:20 UT on 2018 July 5 lasting for 5.8 hours when Algol A passed in front of Algol B. The X-ray spectra of the flare observed with NICER are well reproduced the absorved four-component CIE model, in which two components were fixed to the quiescent-phase parameters the other two for the flare component with its plasma temperature of about 4.15 and 0.90~keV. The peak luminosity at the second MAXI scan is the $5.25^{+0.9}_{-3.8}\times10^{32}~\mathrm{erg~s^{-1}}$, and this is comparable to that of the 1997 event \citep{key-8}. This is one of the largest among the eclipsing X-ray flares. During the eclipse, the 2--10 keV flux decreased by 80\% compared to the out-of-eclipse flux and the temperatures of two CIE components decreased from about 3.96 to 3.26~keV and from 1.15 to 0.14~keV.
  \item We found one example of flare configurations to explain the eclipsing stellar flare, with height and location on the stellar surface are $H=0.8R_{B}$ at 45°S
  using the eclipse timnig in X-rays. The flare loop length and height are also estimated from the peak emission measure $EM$, the peak temperature $T$, and the preflare electron density $n_0$ by using the equation of the magnetic reconnection model in \citet{key-5}. The spetral parametes $EM=4.7\times10^{54}~\mathrm{cm^{-3}}$, $T=4.6\times10^7~\mathrm{K}$, and $n_{0}=10^{10-13}~\mathrm{cm^{-3}}$ give the estimation of $L=(0.2-3.5)\times10^{11}~\mathrm{cm}=0.1-1.4R_{B}$, and $H=(0.1-1.7)\times10^{11}~\mathrm{cm}=0.1-0.7R_{B}$. This flare height is consistent with the values estimated from the eclipse geometry.
  \item The re-increase in the X-ray count, flux, temperature, $EM$, and $N_{\mathrm{H}}$ was observed before the eclipse at MJD 58304.0--58304.4 (Figure~\ref{lc}, \ref{parameta}). This suggests that another flare occurred during the ongoing flare. The simultaneous increase in $N_{\mathrm{H}}$ suggests the possible detection of a CME.
\end{itemize}



\begin{acknowledgments}
We sincerely thank Dr. Jerzy M Kreiner of Mt. Suhora Astronomical Observatory for kindly providing the O-C data \citep{key-4} of Algol used in this study. 
Some/all of the data presented in this article were obtained from the Mikulski Archive for Space Telescopes (MAST) at the Space Telescope Science Institute. The specific observations analyzed can be accessed via \dataset[doi: 10.17909/5dyy-t366]{https://doi.org/10.17909/5dyy-t366}.
The NICER analysis software and data calibration were provided by the NASA NICER mission and the Astrophysics Explorers Program. This research has made use of the MAXI data provided by RIKEN, JAXA and the MAXI team. W.I was supported by JSPS KAKENHI JP24K00673.
T.E. was supported by the JST, Japan grant number JPMJFR202O (Sohatsu). 
Y.N. acknowledges the support from the NASA ADAP award program Number 80NSSC21K0632, the NASA TESS Cycle 6 Program 80NSSC24K0493, and the NASA NICER Cycle 6 Program 80NSSC24K1194.
The work of K.H. is supported by NASA under award number 80GSFC24M0006.
\end{acknowledgments}




%
\facilities{MAXI \citep{key-3}, NICER \citep{key-14}, TESS \citep{key-32}}

\software{astropy \citep{key-43, key-42, key-47}, HEASoft, Xspec \citep{key-12}, SKIRT \citep{key-34, key-49}
          }


\appendix
To verify whether the observed spectrum during the eclipse can be reproduced by these configurations, we tried 3D radiation transfer simulations with the Stellar Kinematics Including Radiative Transfer (SKIRT) toolkit \citep{key-34, key-49}. It is a Monte Carlo simulation to emulate scattering, absorption, and emission by the surrounding medium around the X-ray source. Using this, we tried to simulate the observed eclipsing spectrum by assuming the incident non-eclipsing spectral energy distribution (SED).
The stellar positions of the Algol binary system are
assumed to be at mid-eclipse (Figure~\ref{flare} panels b1, b2). We approximate the emission region by placing three point sources at the base, middle, and top of the flare loop in the cylinder shape, with each emitting intensity at one-third of the total non-eclipse incident X-ray flux. 
As the incident SED setup, we assumed the NICER best-fit spectral model observed before and after the eclipse (MJD 58304.34 and 58304.67, ObsID 1200260102). 
Under these assumptions, we confirmed that the normalization of the eclipsing spectra is reduced by 70\% from the incident model in both configurations. 
The input model assuming the post-eclipse showed a better match to the observed data than the model of the pre-eclipse case. The simulated spectrum assuming the post-eclipse model as the input stays within 0.8--1.2 level of the observed spectral shape in the 0.3--3.0 keV. Thus, we confirmed the assumed configuration approximately reproduce the observed spectrum.

\end{document}